\documentstyle[aps,twocolumn,prl,epsfig]{revtex}

\begin{document}
\draft

\twocolumn[\hsize\textwidth\columnwidth\hsize\csname
@twocolumnfalse\endcsname

\title{Hole spectral functions of LaMnO$_3$}
\author{Wei-Guo Yin$^{1,2}$, Hai-Qing Lin$^{1}$, and Chang-De Gong$^{2}$}
\address{$^1$Department of Physics, The Chinese University of Hong Kong, Sha Tin, Hong Kong, People's Republic of China}
\address{$^2$National Key Laboratory of Solid States of Microstructure, Nanjing University, Nanjing 210093, People's Republic of China}
\date{November 24, 2000}
\maketitle

\begin{abstract}
By use of the orbital $t$-$J$ model, we calculate the photoemission spectra
of LaMnO$_3$ using the exact diagonalization technique, and
interpret our numerics quite well in the orbital-polaron scenario where the
scattering between holes and orbital excitations is treated within the
self-consistent Born approximation.
The quasiparticle bandwidth is found to
be of the order of $J$ and $t$ in the purely Coulombic and Jahn-Teller phononic
model, respectively.
We suggest that angle-resolved photoemission spectroscopy experiments allow one
to distinguish between the orbital-polaron scenario and the
Jahn-Teller polaron scenario.
\end{abstract}


]

Orbital Physics is a key concept for recent intensive studies on
transition-metal oxides \cite{Tokura}, especially on manganese oxides
with perovskite structure $R_{1-x}A_{x}$MnO$_{3}$ ($R$ = La, Pr, Nd, Sm and $%
A$ = Ca, Sr, Ba) due to the discovery of colossal
magnetoresistance in this class of materials. Mn$^{3+}$ in
$R$MnO$_{3}$ has four $d$ electrons of which three are put into
the $t_{2g}$ orbitals and form an $S=3/2$ localized spin, and the
mobile one occupies one of the $e_{g}$ orbitals ($d_{x^{2}-y^{2}}$
and $d_{3z^{2}-r^{2}}$). This $e_{g}$ orbital degree of freedom
may cause new phenomena through strong coupling with charge, spin,
and lattice dynamics \cite{Nagaosa}. For example, alternating
orbital order was observed
in the ferromagnetic (FM) planes of $A$-type antiferromagnetic (AF) LaMnO$%
_{3}$ \cite{Murakami} and uniform $x^{2}-y^{2}$ orbital order in
$A$-type AF Pr$_{1/2}$Sr$_{1/2}$MnO$_{3}$ \cite{Kawano},
while orbital liquid was proposed to exist in metallic FM
La$_{1-x}$Sr$_{x}$MnO$_{3}$ ($x>0.2$) \cite{Quijada}. 
These discoveries attract
attention to the orbital correlation, dynamics and order-disorder
transition. An essential first step is to understand the motion of
a hole in an orbital ordered system such as the FM planes of
LaMnO$_{3}$ \cite{note}. We address it in this Letter.

Orbital ordering in LaMnO$_{3}$ can be induced by either the
Jahn-Teller (JT) lattice distortion originating from the
degeneracy of the $e_{g}$ orbitals
\cite{Millis,Moreo,Hotta} or the intra-atomic
Coulomb interaction in the $e_{g}$ orbitals \cite
{Hotta,Kugel,Ishihara,Horsch,Bala}. In the latter mechanism,
the Coulomb interaction eliminates doubly occupied sites and
results in the so-called {\em orbital} $t$-$J$ model, a purely
electronic Hamiltonian descrbing the orbital sector of physics of
LaMnO$_{3}$. The model was initially derived by Kugel and Khomskii
\cite{Kugel} and has been recently
studied within mean field theory by Ishihara {\em et al.} \cite{Ishihara}
and via a finite-temperature diagonalization method by
Horsch{\em et al. } \cite{Horsch}. The experimentally
observed orbital ordering in LaMnO$_3$ \cite{Murakami} was
reproduced by use of this model \cite{Horsch}. However,
it can be also attributed to Jahn-Teller polarons \cite{Millis,Bala}. In
this Letter, we take the orbital $t$-$J$ model as a starting point
and discuss the JT effect later.

We first calculate photoemission spectra for the orbital $t$-$J$
model using the exact diagonalization technique (ED). The
experimental counterpart of this problem is angle-resolved
photoemission spectroscopy (ARPES) measurements. ARPES experiments
have not yet been reported in this doping regime. Thus, we perform
here computational experiments to explore unbiased information.
Then, we satisfactorily interpret the outcome of such experiments
in an {\em orbital-polaron} picture where the scattering between
holes and orbital excitations is treated within the
self-consistent Born approximation. Furthermore, we point out that
although the JT lattice distortion leads to a large gap in the
orbital excitation spectrum and thus resists the formation of the
orbital-polaron, the hole can move almost freely through
the orbital-flip process. The quasiparticle bandwidth is of the order of 
$J$ ($ t $) in the purely Coulombic (Jahn-Teller phononic) model, 
respectively. 
Our results indicate that ARPES may provide a possible approach to 
distinguish between the Coulombic scenario and the Jahn-Teller scenario.

The orbital $t$-$J$ model is \cite{Kugel,Ishihara,Horsch}
\begin{eqnarray}
H &=&-\sum_{\langle {\bf ij}\rangle \parallel ab}%
(t_{{\bf ij}}^{ab}\widetilde{d}\,_{%
{\bf i}a}^{\dagger }\widetilde{d}\,_{{\bf j}b}^{{}}+{\rm H.c.})  \nonumber \\
&&+\frac{J}{2}\sum_{\langle {\bf ij}\rangle \parallel }[T_{{\bf i}}^{z}T_{%
{\bf j}}^{z}+3T_{{\bf i}}^{x}T_{{\bf j}}^{x}\mp \sqrt{3}(T_{{\bf i}}^{x}T_{%
{\bf j}}^{z}+T_{{\bf i}}^{z}T_{{\bf j}}^{x})],  \label{model}
\end{eqnarray}
where $\widetilde{d}\,_{{\bf i}a}^{\dagger }=d\,_{{\bf i}a}^{\dagger }(1-n_{%
{\bf i}\overline{a}})$ is the constrained fermion operator for the
$e_{g}$ electron at orbital $a$. $T_{{\bf
i}}^{z}=(\widetilde{d}\,_{{\bf i}\uparrow
}^{\dagger }\widetilde{d}\,_{{\bf i}\uparrow }^{{}}-\widetilde{d}\,_{{\bf i}%
\downarrow }^{\dagger }\widetilde{d}\,_{{\bf i}\downarrow }^{{}})/2$ and $T_{%
{\bf i}}^{x}=(\widetilde{d}\,_{{\bf i}\uparrow }^{\dagger }\widetilde{d}\,_{%
{\bf i}\downarrow }^{{}}+\widetilde{d}\,_{{\bf i}\downarrow }^{\dagger }%
\widetilde{d}\,_{{\bf i}\uparrow }^{{}})/2$ are orbital-pseudospin
operators with $|\uparrow \rangle =d_{x^{2}-y^{2}}$ and
$|\downarrow \rangle =d_{3z^{2}-r^{2}}$. 
The anisotropic transfer matrix elements are 
$t_{\bf ij}^{\uparrow \uparrow }=3t/4$, 
$t_{\bf ij}^{\downarrow \downarrow }=t/4$, and 
$t_{\bf ij}^{\uparrow \downarrow }
=t_{\bf ij}^{\downarrow \uparrow }=\mp \sqrt{3} t/4 $, 
here the $\mp $ sign distinguishes hopping along the $x$ and $y$
directions. The orbital superexchange interaction {\bf $J$}$=t^{2}/(U-${\bf $%
J^{\prime }$}$)$ with $U$ ({\bf $J$}$^{\prime }$) being the
interorbital Coulomb (exchange) integral \cite{Ishihara,Horsch}.
In LaMnO$_{3}$, the realistic parameters are estimated from
photoemission experiments \cite {Ishihara,Saitoh} to be $t\sim
0.72$~eV, $U\sim 5$~eV, {\bf $J$}$^{\prime }\sim 2$~eV, thus {\bf
$J$}$\simeq 0.24t$.

The photoemission spectrum (PES) $\langle \widetilde{d}\,_{{\bf k}%
a}^{\dagger }\widetilde{d}\,_{{\bf k}a}^{{}}\rangle _{\omega }$ \cite
{YinPRL} is calculated by using the standard Lanczos algorithm \cite
{Dagotto}. There are two bands and we find $\langle \widetilde{d}\,_{{\bf %
k\downarrow }}^{\dagger }\widetilde{d}\,_{{\bf k\downarrow }}^{{}}\rangle
_{\omega }=\langle \widetilde{d}\,_{{\bf k+Q\uparrow }}^{\dagger }\widetilde{%
d}\,_{{\bf k+Q\uparrow }}^{{}}\rangle _{\omega }$ with ${\bf Q=(}\pi ,\pi
{\bf )}$. Fig. 1 shows the PES $\langle \widetilde{d}\,_{{\bf k\downarrow }%
}^{\dagger }\widetilde{d}\,_{{\bf k\downarrow }}^{{}}\rangle
_{\omega }$ of the ground state of the 16-site cluster at half
filling. At any ${\bf k}$, there is a well-defined quasiparticle
pole (i.e., zero orbital-wave) at the low energy side which is
well separated from a broad, incoherent, multiple-orbital-wave
background extending to the full free-electron
bandwidth. The bottom and the top of the quasiparticle (QP) band locate at $%
(0,0)$ and $(\pi ,\pi )$, respectively. The QP bandwidth of the order of $J$ is
much more narrow than the free-electron bandwidth. Note that all momenta
belonging to the noninteracting 2D Fermi surface $\cos k_{x}+\cos k_{y}=0$
are no longer degenerate despite their close proximity in energy.
Correspondingly, the van Hove singularity in the density of states is
weakened in the orbital $t$-$J$ model.

Since the mixed terms $\propto T_{{\bf i}}^{x}T_{{\bf j}}^{z}+T_{{\bf i}%
}^{z}T_{{\bf j}}^{x}$ in the Hamiltonian (\ref{model}) could contribute only
in higher order orbital-wave theory than the linear one, it is interesting
to perform similar calculations using the ED technique for the simplified
Hamiltonian without these terms, which has been referred to as ``the
truncated Hamiltonian'' by Brink {\em et al.} in their study of the purely
orbital $J$ model \cite{Horsch}. This could let us investigate the
contribution of higher order orbital excitations to the QP behavior. Fig. 1
also illustrates the PES for the truncated Hamiltonian. The overall line
shape of the PES remains almost unchanged but a slight shift to a higher
energy. The prominent influence of higher order orbital excitations is that
they suppress the heights of the quasiparticle peaks and thus weaken the QP
spectral weights. We therefore conclude that as far as the QP dispersion
relation is concerned, higher order orbital excitations can be neglected.
The resultant QP dispersions are displayed in Fig. 2.

To get the physical insight of the above numerics, we perform an
analytical calculation of the QP spectral functions in an {\em
orbital-polaron} scenario. The $J$ term induces the orbital
N\'{e}el state, i.e., the alternation of orthogonal orbitals in
the ground state. Here we analogize our problem to that of a hole
moving in the AF spin background, an essential
problem in the field of high-temperature superconductivity
\cite{SVR}. The latter can be accurately understood in the
spin-polaron scenario where holes are described as spinless
fermions (holons) and spins as hard-core bosons. The scattering
between holes and spin-wave excitations is treated within the
self-consistent Born approximation (SCBA) \cite {SVR}.
In the SCBA, higher order spin-wave excitations are also
neglected, consistent with our ED results. Hence, we expect that
our numerics may be interpreted in a similar way.

In LaMnO$_{3}$, alternatingly occupied orbitals on the two sublattices are
oriented in the FM planes: $(|\uparrow \rangle +|\downarrow \rangle )/\sqrt{2%
}$ and $(|\uparrow \rangle -|\downarrow \rangle )/\sqrt{2}$ \cite
{Murakami,Horsch}. This state is different from the alternating directional
orbitals $3x^{2}-r^{2}$ and $3y^{2}-r^{2}$, which have been naively expected
\cite{Ishihara}. To obtain the N\'{e}el configuration $|T_{{\bf i}}^{z}T_{%
{\bf i}+1}^{z}T_{{\bf i}+2}^{z}T_{{\bf i}+3}^{z}\cdots \rangle =|\downarrow
\uparrow \downarrow \uparrow \cdots \rangle $ in a new orbital basis, we
perform a uniform rotation of orbitals by $90^{\circ }$ about the $T^{y}$
axis: $\widetilde{d}_{{\bf i}\uparrow }\rightarrow (\widetilde{d}_{{\bf i}%
\uparrow }-\widetilde{d}_{{\bf i}\downarrow })/\sqrt{2}$, $\widetilde{d}_{%
{\bf i}\downarrow }\rightarrow (\widetilde{d}_{{\bf i}\uparrow }+\widetilde{d%
}_{{\bf i}\downarrow })/\sqrt{2}$,$\,$and thus $T_{{\bf
i}}^{z}\rightarrow -T\,_{{\bf i}}^{x}$, $T_{{\bf
i}}^{x}\rightarrow T_{{\bf i}}^{z}$. Note that in the new orbital
basis, the interorbital transfer matrix element is changed to
$(t_{\bf ij}^{\uparrow \uparrow }-t_{\bf ij}^{\downarrow \downarrow
})/2$. It is the imbalance between the
$d_{x^{2}-y^{2}}-d_{x^{2}-y^{2}}$ hopping and the
$d_{3z^{2}-r^{2}}-d_{3z^{2}-r^{2}}$ hopping that gives rise to the
interorbital hopping in the rotated basis and causes the bare hole
dispersion as shown below. Orbital excitations can be described
within the
linear spin-wave theory by defining $T\,_{{\bf i}}^{+}=%
\overline{a}_{{\bf i}}^{\dagger }$, $T\,_{{\bf i}}^{z}=\overline{a}_{{\bf i}%
}^{\dagger }\overline{a}_{{\bf i}}^{{}}-\frac{1}{2}$ on the $\downarrow $
sublattice, and $T\,_{{\bf j}}^{+}=\overline{b}\,_{{\bf j}}^{{}}$, $T\,_{%
{\bf j}}^{z}=\frac{1}{2}-\overline{b}\,_{{\bf j}}^{\dagger }\overline{b}\,_{%
{\bf j}}^{{}}$ on the $\uparrow $ sublattice \cite{Horsch}. 
Here $\overline{a}_{{\bf i}}$ and $\overline{b}\,_{{\bf j}}$ are
hard-core boson operators. Considering the orbital N\'{e}el state
as the vacuum state, we define holon operators $\overline{h}_{{\bf
i}}$ ($\overline{f}_{{\bf i}}$) similar to those in the
spin-polaron scenario \cite{SVR} so that
$\widetilde{d}_{{\bf i}\uparrow }=\overline{h}\,_{{\bf
i}}^{\dagger
}\overline{a}_{{\bf i}}^{{}},\,\,\widetilde{\,d}_{{\bf i}\downarrow }=%
\overline{h}\,_{{\bf i}}^{\dagger }$ on the $\downarrow $ sublattice and $%
\widetilde{\,d}_{{\bf j}\downarrow }=\overline{f}\,_{{\bf j}}^{\dagger }%
\overline{b}_{{\bf j}}^{{}},\,\,\,\,\widetilde{d}_{{\bf j}\uparrow }=%
\overline{f}\,_{{\bf j}}^{\dagger }$ on the $\uparrow $ sublattice.

Introducing new fermion operators $\{f_{{\bf k}},h_{{\bf k}}\}$ in the
momentum space $\overline{f}_{{\bf k}}=(f_{{\bf k}}+h_{{\bf k}})/\sqrt{2}%
,\,\,\,\,\overline{h}_{{\bf k}}=(f_{{\bf k}}-h_{{\bf
k}})/\sqrt{2}$, we arrive at an effective orbital-polaron
Hamiltonian
\begin{eqnarray}
H_{{\rm eff}} &=&\sum_{{\bf k}}\varepsilon _{{\bf k}}(f\,_{{\bf k}}^{\dagger
}f_{{\bf k}}^{{}}-h\,_{{\bf k}}^{\dagger }h_{{\bf k}}^{{}})+\sum_{{\bf q}%
}(\omega _{{\bf q}}^{-}\alpha _{{\bf q}}^{\dagger }\alpha _{{\bf q}%
}^{{}}+\omega _{{\bf q}}^{+}\beta _{{\bf q}}^{\dagger }\beta _{{\bf q}}^{{}})
\nonumber \\
&&+\sum_{{\bf kq}}(f\,_{{\bf k}}^{\dagger }f_{{\bf k}-{\bf q}}^{{}}-h\,_{%
{\bf k}}^{\dagger }h_{{\bf k}-{\bf q}}^{{}})(g_{{\bf kq}}^{\beta }\beta _{%
{\bf q}}^{{}}+g_{{\bf kq}}^{\alpha }\alpha _{{\bf q}}^{{}})+{\rm H.c.}
\label{polaron} \\
&&+\sum_{{\bf kq}}(h\,_{{\bf k}}^{\dagger }f_{{\bf k}-{\bf q}}^{{}}-f_{{\bf k%
}}^{\dagger }h_{{\bf k}-{\bf q}}^{{}})(\rho _{{\bf kq}}^{\beta }\beta _{{\bf %
q}}^{{}}+\rho _{{\bf kq}}^{\alpha }\alpha _{{\bf q}}^{{}})+{\rm H.c.},
\nonumber
\end{eqnarray}
where the bare hole dispersion is $\varepsilon _{{\bf k}}=-t\gamma _{{\bf k}%
}^{{}}$. The holon-orbital-wave coupling functions are $g_{{\bf kq}}^{\beta
}=\frac{2t}{\sqrt{N}}(\gamma _{{\bf k}}^{{}}v_{{\bf q}}^{+}+\gamma _{{\bf k}-%
{\bf q}}^{{}}u_{{\bf q}}^{+})$, $\rho _{{\bf kq}}^{\beta }=-\frac{\sqrt{3}t}{%
\sqrt{N}}(\eta _{{\bf k}}^{{}}v_{{\bf q}}^{+}-\eta _{{\bf k}-{\bf q}}^{{}}u_{%
{\bf q}}^{+})$, $\rho _{{\bf kq}}^{\alpha }=g_{{\bf k},{\bf q}+{\bf Q}%
}^{\beta }$ and $g_{{\bf kq}}^{\alpha }=$ $\rho _{{\bf k,q}+{\bf Q}}^{\beta
} $ with $\gamma _{{\bf k}}^{{}}=(\cos k_{x}+\cos k_{y})/2$, $\eta _{{\bf k}%
}^{{}}=(\cos k_{x}-\cos k_{y})/2$, and
\begin{eqnarray}
u_{{\bf q}}^{\pm } &=&\{[(A_{{\bf q}}\pm B_{{\bf q}})/\omega _{{\bf q}}^{\pm
}+1]/2\}^{1/2},  \label{u} \\
v_{{\bf q}}^{\pm } &=&-{\rm sgn}(\pm B_{{\bf q}})\{[(A_{{\bf q}}\pm B_{{\bf q%
}})/\omega _{{\bf q}}^{\pm }-1]/2\}^{1/2}.  \label{v}
\end{eqnarray}
Here $A_{{\bf q}}=3${\bf $J$} and $B_{{\bf q}}=${\bf $J$}$\gamma _{{\bf q}%
}^{{}}/2$. The $\alpha _{{\bf q}}$'s ($\beta _{{\bf q}}$'s) are orbital-wave
operators, $\overline{a}_{{\bf q}}=\frac{1}{\sqrt{2}}(u_{{\bf q}}^{+}\beta _{%
{\bf q}}+v_{{\bf q}}^{+}\beta _{-{\bf q}}^{\dagger }+u_{{\bf q}}^{-}\alpha _{%
{\bf q}}^{{}}+v_{{\bf q}}^{-}\alpha _{-{\bf q}}^{\dagger }),\,\,\,\,\,%
\overline{b}_{q}=\frac{1}{\sqrt{2}}(u_{{\bf q}}^{+}\beta _{{\bf q}}+v_{{\bf q%
}}^{+}\beta _{-{\bf q}}^{\dagger }-u_{{\bf q}}^{-}\alpha _{{\bf q}}^{{}}-v_{%
{\bf q}}^{-}\alpha _{-{\bf q}}^{\dagger })$, with dispersion $\omega _{{\bf q%
}}^{\pm }=\sqrt{A_{{\bf q}}(A_{{\bf q}}\pm 2B_{{\bf q}})}$.

The holon Green's functions $G_{f,h}({\bf k},\omega )=[\omega \mp
\varepsilon _{{\bf k}}-\Sigma _{f,h}({\bf k},\omega )]^{-1}$ are calculated
within the SCBA \cite{SVR}. Note that $\Sigma _{f}({\bf k}+{\bf Q}%
,\omega )=\Sigma _{h}({\bf k},\omega )\ $and $G_{f}({\bf k}+{\bf Q},\omega
)=G_{h}({\bf k},\omega )$ with ${\bf Q=(}\pi ,\pi {\bf )}$. The self-energy
is thus of form
\begin{eqnarray}
\Sigma _{f}({\bf k},\omega ) &=&\sum_{{\bf q}}{}^{\prime }[(g_{{\bf kq}%
}^{\beta })^{2}G_{f}({\bf k}-{\bf q},\omega -\omega _{{\bf q}}^{+})
\nonumber \\
&&+(\rho _{{\bf kq}}^{\beta })^{2}G_{f}({\bf k}+{\bf Q}-{\bf q},\omega
-\omega _{{\bf q}}^{+})],  \label{se}
\end{eqnarray}
where $\sum^{\prime }$ means to sum over the first Brillouin zone.

The QP spectral functions $A_{f}({\bf k},\omega )=-\frac{1}{\pi }%
\mathop{\rm Im}%
G_{f}({\bf k},\omega )$ are shown in Fig. 3 \cite{note2}. The QP dispersion $%
E_{{\bf k}}^{f}=\varepsilon _{{\bf k}}+\Sigma _{f}({\bf k},E_{{\bf k}}^{f})$
(see Fig. 2) and the spectral weights $Z({\bf k})=\left[ 1-\frac{\partial
\Sigma _{f}({\bf k},\omega )}{\partial \omega }\right] _{\omega =E_{{\bf k}%
}}^{-1}$ (see Table I). All of the SCBA results are in good agreement with
the ED results, especially with those for the truncated
Hamiltonian as expected. Therefore, the orbital-polaron
scenario may provide a valuable scheme for further works on the orbital
dynamics. 

In the rest of this Letter, we discuss the Jahn-Teller effect on
the quasiparticle band. Recently, Ba\l a and Ole\'{s} suggested
that a large gap in the orbital excitation spectrum induced by the
JT lattice distortion might lead to a strong confinement of holes
in lightly doped LaMnO$_{3}$ insulators \cite{Bala}. This implies
that the hole quasiparticle bandwidth is more narrow in the
presence of JT phonons. We include the JT effect in the same way
as Ba\l a and Ole\'{s} \cite{Millis,Bala} but we get a
different result. The Jahn-Teller interaction considered here is
\cite {Bala}
\begin{equation}
H_{{\rm JT}}=-2E_{{\rm JT}}(\phi ,\delta _{x},\delta _{z},u)(\sum_{{\bf i}%
\in A}T\,_{{\bf i}}^{z}-\sum_{{\bf j}\in B}T\,_{{\bf j}}^{z}).  \label{JT}
\end{equation}
Here $T\,_{{\bf i}}^{z}$ refers to the rotated basis. $E_{{\rm
JT}}(\phi
,\delta _{x},\delta _{z},u)=\lambda [(\delta _{x}-\delta _{z})\sin 2\phi -2%
\sqrt{3}u\cos 2\phi ]$ acts as a fictitious ``magnetic field'' in which $%
\phi $ is the tilting angle of pseudospins, and $\delta _{x}$, $\delta _{z}$%
, $u$ characterize lattice distortions (in units of the lattice constant): $%
\delta _{x}$($\delta _{z}$) --- uniform deformation along the $x$ and $y$ ($%
z $) directions, and $u$ --- oxygen ionic displacement along Mn-O-Mn bond in
the $xy$ plane \cite{note3}. The distorted lattice energy per Mn ion is $%
E_{l}(\delta _{x},\delta _{z},u)=K_{1}(\frac{1}{2}\delta _{x}^{2}+2u^{2}+%
\frac{1}{4}\delta _{z}^{2})+K_{2}(\delta
_{x}^{2}+\frac{1}{2}\delta _{z}^{2}) $ where $K_{1}(K_{2})$ is the
nearest-neighbor Mn-O (Mn-Mn) spring constant. To estimate the JT
effect on the quasiparticle behavior, we may consider the case of
$\phi =0$ (i.e., no static distortions due to a tetragonal field)
without loss of generality. In the classical ground state at $\phi
=0$, $E_{{\rm JT}}=-3\lambda ^{2}/K_{1}$. For the realistic
parameters of LaMnO$_{3}$: the spring constant $K_{1}=200$~eV and
the JT interaction parameter $\lambda \simeq 6$~eV \cite{Millis,Bala}, $%
E_{{\rm JT}}=-0.54$~eV.

In order to calculate $E_{{\bf k}}^{f}$ in the presence of JT
lattice distortion, all we have to modify the above derivation is
to make the following replacement in (\ref{u}) and (\ref{v}):
$A_{{\bf q}}\rightarrow 3J-2E_{{\rm JT}}=3J+6\lambda ^{2}/K_{1}$.
The JT interaction adds an Ising-like component to the excitations
and induces a large gap in the orbital excitation spectrum. Thus,
the JT effect stabilizes the orbital ordering.

The SCBA results on the $20\times 20$ lattice are summarized in
Table 1. The large gap in the orbital excitation spectrum induced
by the JT lattice distortion does not weaken the QP bandwidth.
Instead, at $\lambda =6$ eV the QP bandwidth reaches $75$ percent 
of the width ($2t$) of the bare hole dispersion $\varepsilon _{{\bf k}}$.
Although the large gap in the orbital excitation spectrum
resists the formation of the orbital polaron, the hole can move almost {\em %
freely} via the orbital-flip process in the rotated orbital basis. For $A_{%
{\bf q}}\gg t$ because of large $J$ or large $\lambda $, Eq. (\ref{se}) can
be solved analytically in perturbation theory. Most of the spectral weight, $%
1-O(t^{2}/A_{{\bf q}}^{2})$, appears in the quasiparticle part of $G_{f}(%
{\bf k},\omega )$\thinspace which behaves indeed a bare dispersion $E_{{\bf k%
}}^{f}\simeq \varepsilon _{{\bf k}}-\sum_{{\bf q}}^{\prime }[(g_{{\bf kq}%
}^{\beta })^{2}+(\rho _{{\bf kq}}^{\beta })^{2}]/\omega _{{\bf
q}}^{+}\simeq \varepsilon _{{\bf k}}-O(t^{2}/A_{{\bf q}})$. 
Fig. 4 shows the QP bandwidth $W$ as a function of $J$.
Without the JT interaction ($\lambda=0$~eV), 
$W\simeq 2.2J$ scales with $J$ in the range of $0.01\leq J\leq 0.4$, 
and approaches to the width of the bare hole dispersion ($2t$) for large $J$.
On the other hand, with strong JT interaction ($\lambda=6$~eV), 
$W$ is of the order of $t$ in the whole range of $J$. 
These results reminisce those for one hole moving in
the {\em spin} $t$-$t^{\prime }$-$J$ model 
(the $t$-$t^{\prime }$-$J_z$ model with large $J_z$),
respectively, where the transfer to next nearest neighbors
($t^{\prime }$) provides the bare hole dispersion \cite {YinPRL,Dagotto}.
Therefore, the QP bandwidth for realistic $J$ is of the order of 
$t$ ($J$) in the presence (absence) of strong Jahn-Teller
interaction. 
Our results can be tested by future ARPES experiments.

The authors thank P. W. Leung, F. C. Zhang, and H. Zheng for
useful discussions. The ED program was implemented on the basis of $C$
codes provided by P. W. Leung. W.G.Y. is supported by CUHK 4288/00P 2160148.

\vskip 1cm
\begin{table}[btp]
\caption{Bandwidth $W$ and spectral weights $Z(k)$ as a function of the
Jahn-Teller interaction parameter $\lambda$. $J=0.3t$.}
\label{table 2}
\begin{tabular}{cccccc}
$\lambda$ (eV) & $W/t$ & $Z(0,0)$ & $Z(\pi/2,\pi/2)$ & $Z(\pi,\pi)$ & $%
Z(\pi,0)$ \\
\tableline 0 & 0.65210 & 0.65989 & 0.37446 & 0.09774 & 0.39068 \\
4 & 1.12691 & 0.79211 & 0.60401 & 0.30239 & 0.60701 \\
6 & 1.49980 & 0.87222 & 0.77294 & 0.57996 & 0.77022 \\
8 & 1.74565 & 0.92421 & 0.88066 & 0.80063 & 0.87943
\end{tabular}
\end{table}


\begin{figure}[tbp]
\begin{center}
\epsfig{file=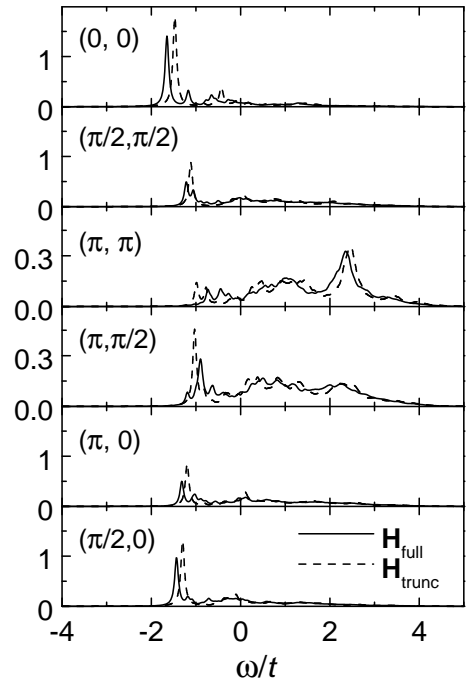,width=0.9\hsize,angle=0}
\end{center}
\vskip -0.5cm
\caption{The PES $\langle \widetilde{d}\,_{{\bf k}\downarrow }^{\dagger }%
\widetilde{d}\,_{{\bf k}\downarrow }^{}\rangle _{\omega }$ for the orbital $%
t $-$J$ model on the $4\times 4$ cluster using the ED technique for the full
Hamiltonian (solid lines) and the truncated Hamiltonian (dashed lines). Here
$J=0.3t$.}
\label{fig. 1}
\end{figure}

\begin{figure}[tbp]
\begin{center}
\epsfig{file=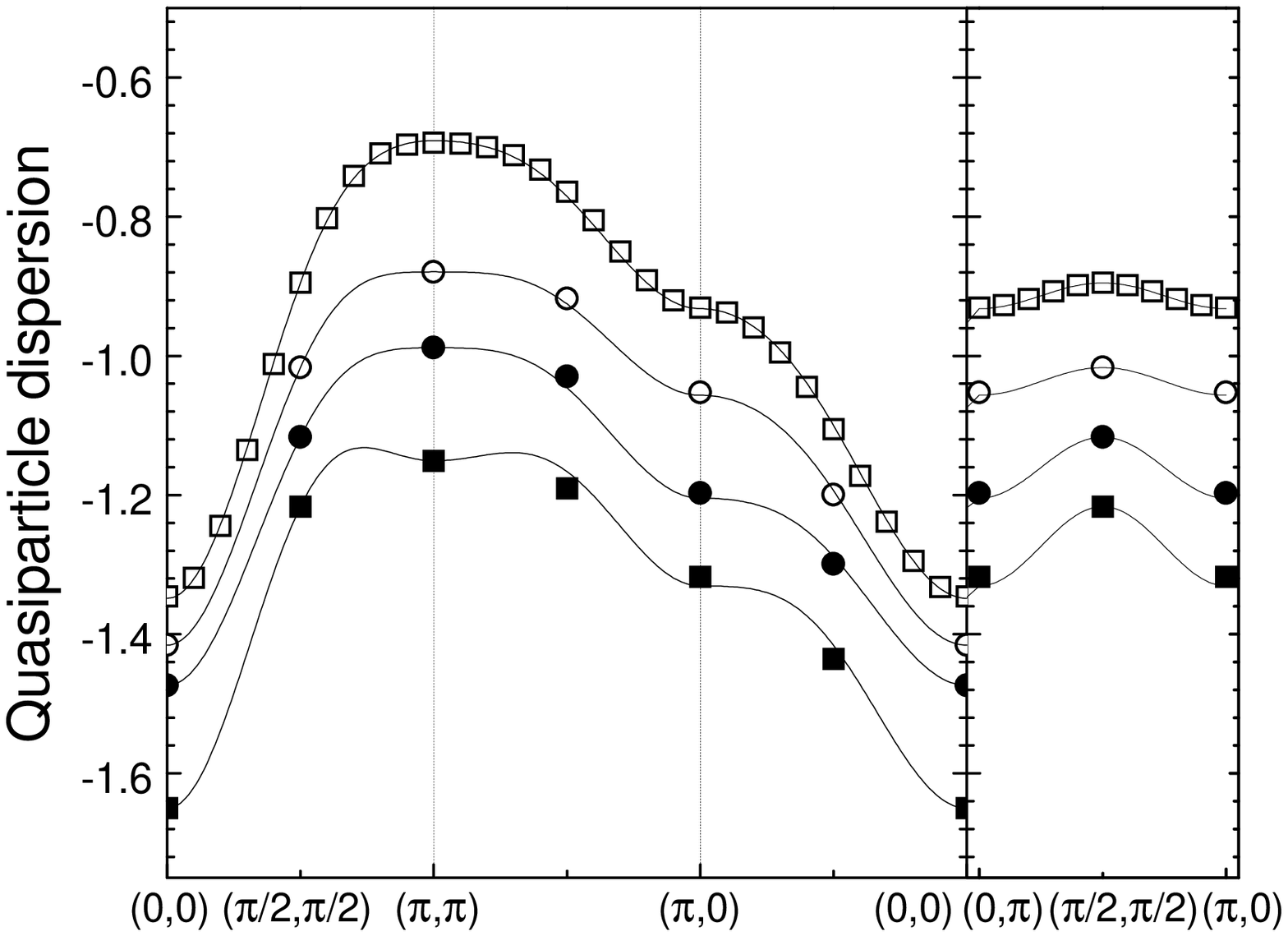,width=1.0\hsize,angle=0}
\end{center}
\vskip -5.5cm %
\caption{Hole quasiparticle dispersion for the orbital $t$-$J$
model with $J=0.3t$: The ED results for the full Hamiltonian
(solid squares) and the
truncated Hamiltonian (solid circles); The SCBA results on the $%
4\times 4$ cluster (open circles) and the $20\times 20$ one (open
squares). Solid lines are the fits using $E_{{\bf k}%
}^{f}=a_{0}+a_{1}\gamma _{{\bf k}}+a_{2}\cos k_{x}\cos k_{y}+a_{3}\gamma _{2%
{\bf k}}$.}
\label{fig. 2}
\end{figure}

\begin{figure}[tbp]
\begin{center}
\epsfig{file=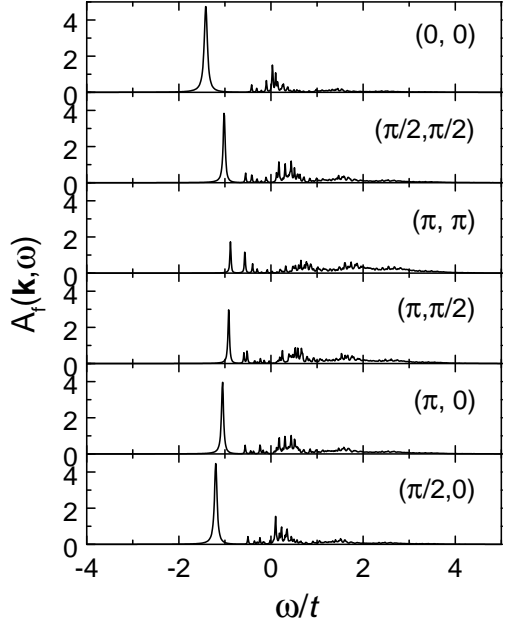,width=0.9\hsize,angle=0}
\end{center}
\vskip -1cm %
\caption{The PES $A_f({\bf k},\omega)$ for the orbital $t$-$J$
model on the $4\times 4$ cluster calculated within the SCBA. Here
$J=0.3t$.} \label{fig. 3}
\end{figure}

\begin{figure}[tbp]
\begin{center}
\epsfig{file=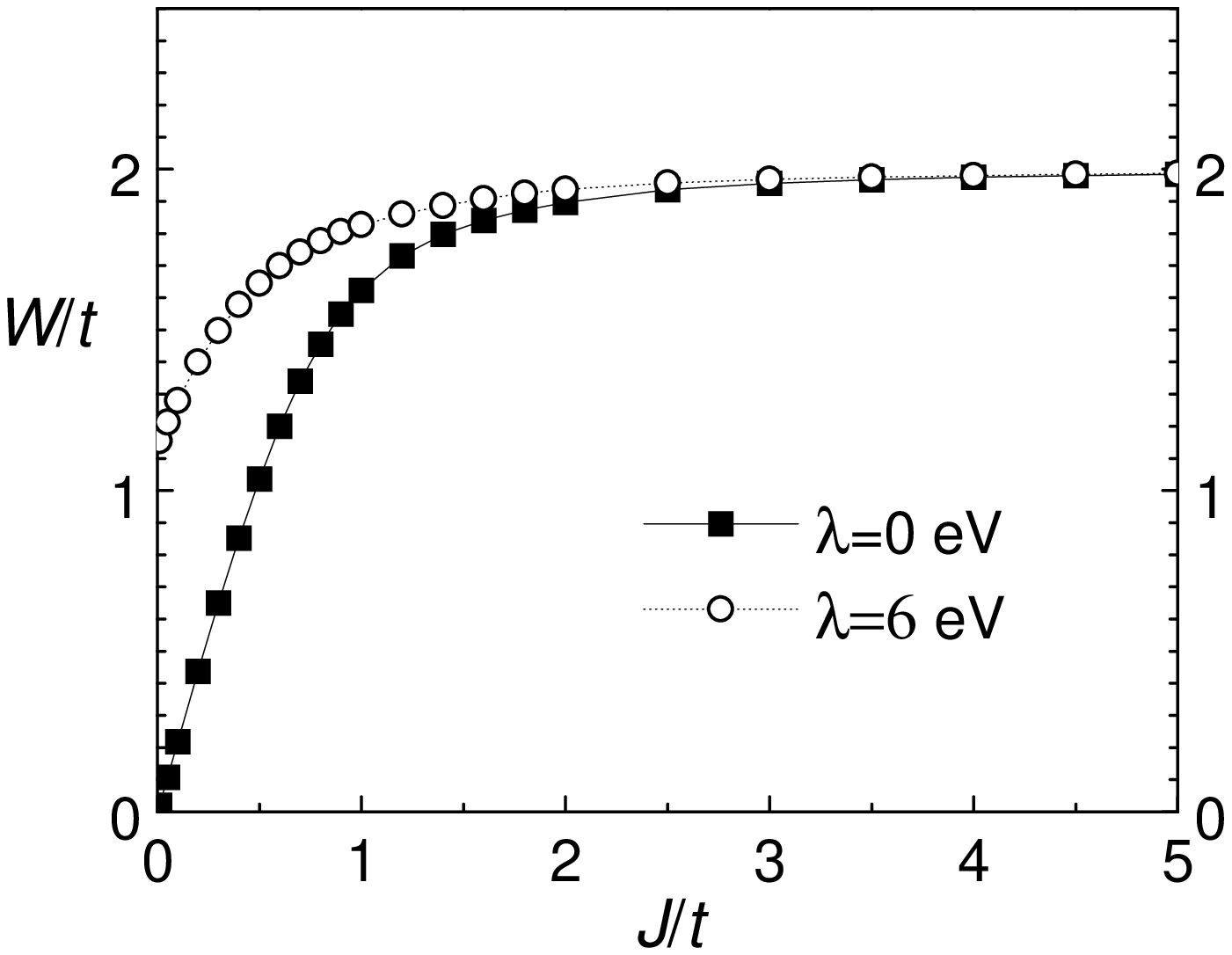,width=1.0\hsize,angle=0}
\end{center}
\vskip -6cm %
\caption{The quasiparticle bandwidth $W$ as a
function of $J$ at different Jahn-Teller interaction parameters:
$\lambda=0$ (solid squares) and $\lambda=6$ eV (open circles).
Calculations are performed on the $16\times 16$ cluster within the
SCBA.} \label{fig. 4}
\end{figure}


\begin{references}
\bibitem{Tokura}  Y. Tokura and N. Nagaosa, Science {\bf 288}, 462 (2000);
Y. Q. Li, M. Ma, D. N. Shi, and F. C. Zhang, Phys. Rev. Lett. {\bf 81}, 3527 (1998).

\bibitem{Nagaosa}  R. Maezono, S. Ishihara, N. Nagaosa, Phys. Rev. B {\bf 58}%
, 11 583 (1998).

\bibitem{Murakami}  Y. Murakami, J. P. Hill, D. Gibbs, M. Blume, I. Koyama,
M. Tanaka, H. Kawata, T. Arima, Y. Tokura, and Y. Endoh, Phys. Rev. Lett.
{\bf 81}, 582 (1998).

\bibitem{Kawano}  H. Kawano, R. Kajimoto, H. Yoshizawa, Y. Tomioka, H.
Kuwahara, and Y. Tokura, Phys. Rev. Lett. {\bf 78}, 4253 (1997).

\bibitem{Quijada}  
K. H. Kim, J. H. Jung, and T. W. Noh, Phys. Rev. Lett. {\bf 81}, 1517 (1998);
M. Quijada, J. \"{C}erne, J. R. Simpson, H. D. Drew, K.
H. Ahn, A. J. Millis, R. Shreekala, R. Ramesh, M. Rajeswari, and T.
Venkatesan, Phys. Rev. B {\bf 58}, 16 093 (1998).

\bibitem{note}  Because of Hund's rule, the spin of the mobile $e_{g}$
electron is enforced to be parallel to the localized $t_{2g}$ electrons, and
the transfer integrals are normalized via the double-exchange mechanism [C.
Zener, Phys. Rev. {\bf 82}, 403 (1951); P. W. Anderson and H. Hasegawa,
Phys. Rev. {\bf 100}, 675 (1955)] by a factor of $\cos (\theta _{xy}/2)$ or $%
\cos (\theta _{z}/2)$ where $\theta _{xy}$, $\theta _{z}$ are the angle
between neighboring $t_{2g}$ spins in the $xy$ plane and in the $z$%
-direction, respectively. Thus, hole motion in the perfect
$A$-type AF state ($\theta _{xy}=0$ and $\theta _{z}=\pi $) can be
considered as hole motion in the FM planes.


\bibitem{Millis}  A. J. Millis, Phys.\ Rev. B {\bf 53}, 8434 (1996);
K. H. Ahn and A. J. Millis, {\em ibid.} {\bf 58}, 3697 (1998);
J. Kanamori, J. Appl. Phys. {\bf 31}, 145 (1961).

\bibitem{Moreo}  
T. Hotta, S. Yunoki, M. Mayr, and E. Dagotto, Phys. Rev. B 
{\bf 60}, R15 009 (1999);
A. Moreo, S. Yunoki, and E. Dagotto, Science {\bf 283}, 2034 (1999);
S. Yunoki, A. Moreo, and E. Dagotto, Phys.\ Rev. Lett. {\bf 81}, 5612 (1998).

\bibitem{Hotta}  T. Hotta, A. L. Malvezzi, and E. Dagotto, Phys.\ Rev. B
{\bf 62}, 9432 (2000).

\bibitem{Kugel}  K. I. Kugel and D. I. Khomskii, Zh. \'{E}ksp. Teor. Fiz.
{\bf 64}, 1429 (1973) [Sov. Phys. JETP {\bf 37}, 725 (1973)].

\bibitem{Ishihara}  S. Ishihara, J. Inoue, and S. Maekawa, Phys. Rev. B {\bf %
55}, 8280 (1997);
S. Ishihara, M. Yamanaka, and N. Nagaosa, {\em ibid.} {\bf 56}, 686 (1997);
R. Kilian and G. Khaliullin, {\em ibid.} {\bf 58}, R11
841 (1998).

\bibitem{Horsch}  P. Horsch, J. Jaklic, and F. Mack, Phys. Rev. B {\bf 59},
6217 (1999);
J. van den Brink, P. Horsch, and F. Mack, {\em ibid.} {\bf 59}, 6795 (1999);
F. Mack and P. Horsch, Phys. Rev. Lett. {\bf 82}, 3160
(1999).

\bibitem{Bala}  J. Ba\l a and A. M. Ole\'{s}, Phys. Rev. B {\bf 62}, R6085
(2000).

\bibitem{Saitoh}  T. Saitoh, A. E. Bocquet, T. Mizokawa, H. Namatame, A.
Fujimori, M. Abbate, Y. Takeda, and M. Takano, Phys. Rev. B {\bf 51}, 13 942
(1995).

\bibitem{YinPRL}  Wei-Guo Yin, C. D. Gong, and P. W. Leung, Phys. Rev. Lett.
{\bf 81,} 2534 (1998);
P. W. Leung and R. J. Gooding, Phys. Rev. B {\bf 52}, R15
711 (1995).

\bibitem{Dagotto}  E. Dagotto, Rev. Mod. Phys. {\bf 66}, 763 (1994).

\bibitem{SVR}  S. Schmitt-Rink, C. M. Varma, and A. E. Ruckenstein, Phys.
Rev. Lett. {\bf 60}, 2793 (1988); 
C. L. Kane, P. A. Lee, and N. Read, Phys. Rev. B {\bf 39}, 6880 (1989);
G. Martinez and P. Horsch, {\em ibid.} {\bf 44}, 317 (1991);
Z.Liu and E. Manousakis, {\em ibid.} {\bf 44}, 2414 (1991);
Wei-Guo Yin, Biao Hao, and C. D. Gong, Phys. Lett. A {\bf 220}, 281 (1996).

\bibitem{note2}  In comparison with the ED results, we perform the following
transformation for the SCBA results: $\omega \rightarrow \omega +\Delta
E_{J} $ where $\Delta E_{J}=0.43903t$ is the energy loss of the orbital
background at one hole doping.

\bibitem{note3}  When a hole is present on site $i$, it attracts the
surrounding oxygen ions equally, giving rise to a breathing distortion
energy. Since the breathing mode does not distinguish between the $e_{g}$
orbitals, it can be safely neglected in the present study.
\end{references}
\end{document}